# Evaluation of the Effects of Compressive Spectrum Sensing Parameters on Primary User Behavior Estimation

*Ahmed A. Tawfik[1], Mohamed F. Abdelkader[2], Sherif M. Abuelenin[3]*

## ABSTRACT

As the Internet of Things (IoT) technology is being deployed, the demand for radio spectrum is increasing. Cognitive radio (CR) is one of the most promising solutions to allow opportunistic spectrum access for IoT secondary users through utilizing spectrum holes resulting from the underutilization of frequency spectrum. A CR needs to frequently sense the spectrum to avoid interference with primary users (PUs). Compressive spectrum sensing techniques have been gaining increasing interest in wideband spectrum sensing, as they reduce the need for high-rate analog-to-digital converters, reducing the complexity and energy requirements of the CR. To enhance spectrum sensing performance, researchers proposed to incorporate PU spectrum usage information into the process of spectrum sensing. Spectrum usage information can be obtained through pilot signals, geo-locational databases or through evaluation of previous spectrum sensing results. In this paper, we are studying the effects of compressive sensing parameters namely compression ratio, sensing period, and sensing duration on the estimation of primary user behavior statistics. We achieved an accurate estimation of the primary user's behavior while saving 40% of the sampling rate by using compressive spectrum sensing compared to traditional spectrum sensing with Nyquist rate sampling.
**Keywords**:   Cognitive radio; Compressive spectrum sensing; Primary user activity statistics

## 1. INTRODUCTION

The demand for frequency spectrum has been increasing in the past few years and is expected to increase more in the near future as wireless technologies such as Internet of Things (IoT) are being deployed [1]. On the other hand, research studies [2] showed that the frequency spectrum has been underutilized as primary users (PUs) or licensed users are not utilizing their channels the whole time. This can allow secondary users (SUs) or unlicensed users to communicate using spectrum holes left unused by primary users. One of the most promising solutions to allow opportunistic spectrum access for IoT devices is Cognitive Radio (CR) [3], which was defined by Haykin in [4] as an intelligent digital radio that understands the surrounding environment and adapts the radio parameters according to its understanding to achieve a reliable communication and efficient use of the radio spectrum.

For a CR unit to understand its environment, it has to frequently sense the spectrum to detect spectrum holes which can be used for transmission while PU is absent. To help CR systems achieve better performance, researchers proposed different methods to exploit the knowledge of PU behavior to better choose channels for transmission, reduce interference with PUs, facilitate spectrum management and provide efficient use of resources. One approach was using a pilot signal to broadcast real-time information about the spectrum utilization and primary users' presence which could be received by secondary users and used to enhance SU spectrum access and avoid interference [5]. However, this method requires physical adjustments to the primary system which may not be possible for some systems or requires additional cost to the existing PU system. In [6] and [7] authors proposed using a geolocation database to provide SUs with regional information about PU system location, transmitted power, vacant channels, etc. However, this solution requires continuous update of database information and SUs must periodically communicate with these databases which results in communication overhead.

Another solution for SUs to acquire information about PU behavior statistics is to estimate this information through spectrum sensing. In [8], the authors proposed to estimate the average duty cycle of a single frequency channel utilizing the prior knowledge of mean PU idle and busy times. On the other hand, in [9] the authors used Fisher Information to measure the performance of using uniform and random sampling to sense the spectrum and to estimate the mean idle time of a PU. They found that random sampling provides better accuracy than uniform sampling for a fixed sensing window. However, in both[8] and [9] the idle and busy times were assumed to follow exponential distribution that has been proved in [10] and [11] to be an imperfect fit in practical scenarios.

Authors in [12] proposed a weighted spectrum sensing technique combined with a machine learning approach to estimate the actual occupancy of the spectrum by exploiting cooperative spectrum sensing. Neighboring

_________________________________________________

[1] *Electrical Engineering Department, Faculty of Engineering, Port Said University, Port Said, Egypt, email: a.abdelkareem@eng.psu.edu.eg, Corresponding author*

[2] *Electrical Engineering Department, Faculty of Engineering, Port Said University, Port Said, Egypt, email:* mdfarouk@eng.psu.edu.eg

[3] *Electrical Engineering Department, Faculty of Engineering, Port Said University, Port Said, Egypt, email:* s.abuelenin@eng.psu.edu.eg

SUs report their recovered spectrum occupancy to a fusion center which uses a machine learning voting algorithm to decide the actual spectrum occupancy and assigns channels back to SUs.

Based on periodic observations of PU behavior, authors in [13] used Bayesian non-parametric models and goodness of fit techniques to estimate the distributions of PU activity utilizing a cooperative spectrum sensing scheme where CR nodes having similar observation of the PU behavior are grouped together. However, the investigated results were based on a fixed number of primary users. In [14] and [15], the authors proposed a modification to the method of moments estimation technique to estimate the distribution of idle and busy times of a single channel using spectrum sensing decisions. The idle and busy periods were assumed to follow the Generalized Pareto (GP) distribution [16] which was considered a good fit in practical scenarios, and they also provided a hardware experiment to confirm their findings. However, their work was based on a single channel and the spectrum was sensed using energy detection at a high sampling rate.

For a CR unit to utilize spectrum holes and avoid interfering with PUs it needs to frequently sense radio channels. This requires sampling narrowband channels using an analog-to-digital converter (ADC) [17]. For wideband spectrum sensing, a very high sampling rate ADC or a bank of lower rate ADCs are required. This leads to higher energy consumption and increased system complexity.

Compressive sensing [18] has been proposed as a more suitable solution for wideband spectrum sensing [19]. Since the frequency spectrum is underutilized, this means it has a sparse representation [20] in frequency domain which facilitates the exploitation of compressive sensing to sample the spectrum by a sub-Nyquist rate and reconstruct the original spectrum, eliminating the need for expensive high-speed ADC. Authors in [21] used analog-to-information converter (AIC) to reconstruct the PSD of the spectrum by a low number of measurements to detect the presence of primary users. In [22] a two-step compressive sensing approach was introduced. In the first step the sparsity of the signal was estimated, then in the second step the needed number of measurements for signal reconstruction was computed. Authors in [23] studied the gain in performance achieved by using a distributed compressive spectrum sensing approach where a centralized fusion centre receives the autocorrelation of the signal samples from a set of CRs, after that a recovery algorithm is used to recover the spectrum by exploiting the joint sparsity. In [24] authors incorporated PU activity statistics as prior information into two schemes to achieve a better performance in compressive spectrum sensing. In the first scheme, a prediction technique was used to predict the support of the signal at the next time instant to be used to reduce the search space for the recovery algorithm. The second scheme used the duty cycles that represent the spectrum usage to solve a weighted $l_1$ minimization problem which achieved a better performance in spectrum recovery at lower compression ratios. However, the duty cycles of the frequency channels were assumed to be given to the CR.

In this paper, we propose the usage of compressive spectrum sensing decisions to estimate the PU behavior statistics using the modified method of moments proposed by the authors in [14] and [15]. We also study the effects of compressive spectrum sensing parameters namely compression ratio, sensing period, and sensing duration on the accuracy of the estimated PU behavior statistics over a sparse wideband spectrum having channels with different duty cycles.

This paper is organized as follows. In Section 2 we describe the system model used in our simulation. In Section 3 we discuss the process to model and estimate the behavior statistics of a PU system. In Section 4 we study the effect of compressive spectrum sensing parameters on the estimation of PU behavior statistics based on the modified method of moments estimation discussed in Section 3. And finally, we conclude the paper in Section 5.

## 2. SYSTEM MODEL

In this work we consider a CR system consisting of a single SU sensing a sparse wideband spectrum of non-overlapping frequency channels. The number of active PUs is random and the channel between each PU and the SU is an Additive White Gaussian Noise (AWGN) channel while the power level of the PU in each channel is assumed to be unknown. For $L$ active PUs, the signal transmitted by each PU is denoted $si(t)$ where $i \in [1, L]$. The signal received by the SU terminal can be represented as

$$x(t) = \sum_{i=1}^{L} s_i(t) + w(t) \quad (1)$$

where $w(t)$ is the AWGN.

By applying discrete Fourier transform (DFT) to eq. (1) we can obtain the frequency spectrum of the received signal as

$$X = S + W \quad (2)$$

where $S$ denotes the transmitted signal and $X$ is an $N \times 1$ vector representing the received spectrum at the SU.

In order for the SU to recover the spectrum $S$, the SU needs to sample the received signal. To achieve that, an AIC could be used to sample the signal with a low number of samples compared to the usage of an ADC, which requires the signal to be sampled at Nyquist rate. Assuming a sparse spectrum where the number of active channels $L$ is much smaller than the total number of channels $N$, the AIC can sample spectrum with a compression ratio of $(M/N)$, which is defined as the number of compressed measurements divided by the total length of the sparse signal, to collect $M$ measurements. The measurement samples vector $y$ can be represented as $M \times 1$ vector where $(L < M << N)$ as follows

$$y = \Phi x = \Phi F^{-1} X \quad (3)$$

where $\Phi$ is the $M \times N$ measurement matrix and $F^{-1}$ is the $N \times N$ inverse DFT matrix. To ensure perfect recovery of the sparse signal the measurement or sensing matrix $\Phi$ must satisfy the restricted isometry property (RIP) [25] condition which states that if there exists an isometry constant $\delta_s$ where $0 < \delta_s < 1$ it must satisfy the following inequality [25].

$$(1 - \delta_s)\|x\|_{l2}^2 \leq \|\Phi x\|_{l2}^2 \leq (1 + \delta_s)\|x\|_{l2}^2 \quad (4)$$

where $\|.\|_{l2}$ represents the $l_2$-norm.

This can be achieved by selecting the measurement matrix to be a Gaussian matrix and the number of measurements must be in the order of $C\,L\log(N/L)$ where $C$ is a constant. As the number of rows $M$ is smaller than the number of columns $N$ of the sensing matrix $\Phi$, there would be an infinite number of solutions for this system of equations, however we are only interested in the sparsest solution. The sparsest solution for this system of equations is the minimization of $\|x\|_0$ which is an NP-hard problem. A more feasible solution is to relax the problem to a $l_1$-norm minimization which tends to converge to the sparsest solution by solving the following convex optimization problem as in [19]

$$\min_x \|x\|_{l1} \text{ Subject to } y = \Phi x \quad (5)$$

This convex optimization problem can be solved by many algorithms such as greedy algorithms like matching pursuit (MP) [26], orthogonal matching pursuit (OMP) [27], and CoSaMP [28], or convex relaxation approaches such as Basis Pursuit (BP) [29] and Basis Pursuit Denoising (BPDN) [30].

## 3. ESTIMATION OF PU STATISTICS USING COMPRESSIVE SPECTRUM SENSING

In this paper, we consider a sparse spectrum of $N$ non-overlapping frequency channels. The channels were grouped into $j$ groups and each channel has a duty cycle $\psi_i$ where $\psi_i \in \{\Psi_1, \Psi_2, \Psi_3, ..., \Psi_j\}$. The average duty cycle $\overline{\Psi} = \frac{1}{N}\sum_{j=1}^{N}\psi_j$ is set to a low value to reproduce a sparse domain. To model the behavior of the PU in each channel we used Continuous-Time Semi-Markov Chain (CTSMC) [31], in which the time index is considered to be continuous. According to this model the PU stays in a single state for a random amount of time then changes to the other state for a random amount of time. The state holding times (idle and busy periods) can follow any statistical distribution which differs from Continuous-Time Markov Chain where the idle/busy periods can only follow exponential distribution. Our aim is to estimate the distribution of idle/busy periods through periodic spectrum sensing by utilizing the channel states to estimate the length of idle/busy periods. If the spectrum is sensed by a high sampling rate (high-resolution modeling), this enables accurate characterization of spectrum usage which is appropriate for a short time scale as high sampling rate results in large number of samples which leads to huge memory requirement and high computational processing. On the other hand, monitoring the spectrum with a lower sampling rate (low-resolution modeling) was found to be a more accurate approach to model how a CR would observe the spectrum. In low-resolution modeling idle and busy periods were found to best fit the Generalized Pareto (GP) distribution over a wide range of radio technologies [32]. The cumulative distribution function (CDF) of GP distribution is expressed [32] as follows:

$$F_{GP}(T;\mu,\lambda,\alpha) = 1 - \left[1 + \frac{\alpha(T-\mu)}{\lambda}\right]^{-1/\alpha} \quad (6)$$

where $T$ represents time period length and $\mu, \lambda, \alpha$ are location, scale and shape parameters respectively. The GP parameters satisfy these conditions:

$$T \geq \mu \text{ for } \alpha > 0$$
$$\mu \leq T \leq \mu - \frac{\lambda}{\alpha} \text{ for } \alpha < 0$$
$$\mu > 0, \lambda > 0 \text{ and } \alpha < 1/2$$

The mean $\mathbb{E}\{T\}$ and the variance $\mathbb{V}\{T\}$ are given in [32] as

$$\mathbb{E}\{T\} = \mu + \frac{\lambda}{1-\alpha} \quad (7)$$

$$\mathbb{V}\{T\} = \frac{\lambda^2}{(1-\alpha)^2(1-2\alpha)} \quad (8)$$

Another important parameter to consider when modeling the PU behavior in this work is the duty cycle (DC) which characterizes the spectrum usage and defined as the portion of time that a channel is busy or the probability of finding a channel used by a PU. In (CTSMC) model the duty cycle denoted by $\Psi$ can be expressed in [31] as follows:

$$\Psi = \frac{\mathbb{E}\{T_{busy}\}}{\mathbb{E}\{T_{busy}\} + \mathbb{E}\{T_{idle}\}} \quad (9)$$

where $\mathbb{E}\{T_{busy}\}$ is the mean value of busy periods and $\mathbb{E}\{T_{idle}\}$ is the mean of idle periods.

In our simulations each channel was assigned a sequence of idle and busy periods $T_{i,n}$ where $i = 0$ for idle and 1 for busy and '$n$' represent the number of periods. These periods were chosen to follow GP distribution and the GP distribution parameters were selected to reproduce each duty cycle value using eqs. (7) and (9). To simulate the signal transmitted by PUs, the sequence of the periods in each channel is sampled to binary ones and zeros to represent the channel states, then the ones are multiplied by a random complex number to represent the power of the transmitted signal. The binary samples can be represented in an $N \times V$ matrix, where $N$ is the number of channels and $V$ is the number of samples in a certain sensing duration which refers to how long the spectrum was sensed. The longer the SU monitors the spectrum, the more it can accurately estimate the behavior of the PU. Then random Gaussian noise is added to the samples to model an AWGN channel.

We then consider a SU sensing the spectrum every $T_s$ seconds to estimate the behavior of the PU in each channel. Basis Pursuit Denoising (BPDN), which solves the following convex minimization problem as in [30] was chosen to recover the spectrum through the measurement samples acquired by the SU,

$$\min_x \|x\|_{l1} + \lambda\|y - \Phi x\|_2^2 \quad (10)$$

where $\lambda$ is a penalty parameter that can be estimated if noise variance is known. Each column of the $N \times V$ matrix which represents the spectrum at a given moment of time is multiplied by $M \times N$ sensing matrix to acquire the needed compressed measurements for BPDN to reconstruct the spectrum. The binary spectrum decisions are then used for the estimation of the distribution of idle and busy periods and the duty cycle of each channel.

### 3.1. Periods Reconstruction and Mean and Variance Estimation

The binary decisions obtained from spectrum sensing are utilized for reconstruction of the lengths of the idle and busy periods as shown in Fig. 1. The time between two state changes (denoted by $T_i$) is considered an estimate of the real activity period where $i = 1$ represents a busy period and $i = 0$ represents an idle period. Thus, activity (idle/busy) periods of the PU in each channel can be estimated as integer multiples of the sensing period $T_s$, which is defined as the time between two sensing events (i.e. $\{\hat{T}_{i,n}\} = KT_s$ or $(K+1)T_s$) according to the position where the sensing started with respect to the original period, where '$K$' is an integer representing the number of consecutive binary decisions of the same channel state and '$n$' represents the index of the periods. To model the estimation error, the estimated period $\hat{T}_i$ can be rewritten as

$$\hat{T}_i = \left(\left\lfloor \frac{T_i}{T_s} \right\rfloor + \xi\right) T_s \quad (11)$$

where $\xi \in \{0,1\}$ is a Bernoulli random variable and $\lfloor . \rfloor$ is the floor operator.

As shown in Fig. 1 the actual activity period $T_0$ can be estimated in three ways:
1) ES1: $\hat{T}_0 = t_b - t_a$, where the estimated period would be an underestimation of the actual period.
2) ES2: $\hat{T}_0 = t_y - t_x$, where the estimated period would be an overestimation of the actual period.
3) ES3: $\hat{T}_0 = \frac{(t_b - t_a) + (t_y - t_x)}{2}$, where the estimated period would be a more accurate estimation of the actual period.

The minimum activity period (denoted as $\mu_i$) can be estimated as

$$\hat{\mu}_i = \min(\hat{T}_i) = \left\lfloor \frac{\mu_i}{T_s} \right\rfloor \quad (12)$$

Given a set of $N_p$ periods the mean and variance of the periods can be calculated as

$$\mathbb{E}(\hat{T}_i) = \frac{1}{N_p} \sum_{n=1}^{N_p} \hat{T}_{i,n} \quad (13)$$

$$\mathbb{V}(\hat{T}_i) = \frac{1}{N_p - 1} \sum_{n=1}^{N_p} (\hat{T}_{i,n} - \mathbb{E}(\hat{T}_i))^2 \quad (14)$$

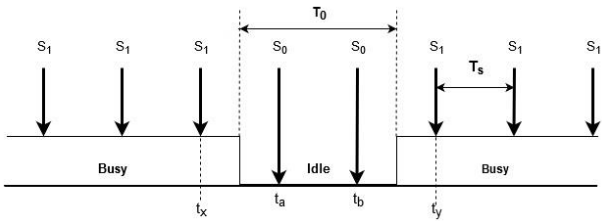

**Figure 1: Channel binary decisions used to estimate activity periods. $S_1, S_0$ represent busy and idle states respectively while $T_s$ represents the sampling period (sensing period), and $T_0$ is an actual idle period.**

### 3.2. Estimation of the Distribution of PU Activity Periods

The Method of Moments (MoM) can be used to estimate the distribution of the PU activity periods accurately when they are supposed to follow a specific distribution. It can also estimate the distribution of the idle and busy periods without the need to store the history of periods' values as the mean and variance can be calculated recursively for each new reconstructed period. According to MoM we can firstly estimate the mean and variance (moments) of the periods following GP distribution as in eqs. (13) and (14). Then the parameters of the GP distribution can be calculated using the moments as follows

$$\hat{\alpha} = \frac{1}{2}\left(1 - \frac{(\mathbb{E}(\hat{T}_i) - \hat{\mu}_i)^2}{\mathbb{V}(\hat{T}_i)}\right) \quad (15)$$

$$\hat{\lambda} = \frac{1}{2}\left(1 + \frac{(\mathbb{E}(\hat{T}_i) - \hat{\mu}_i)^2}{\mathbb{V}(\hat{T}_i)}\right)\left(\mathbb{E}(\hat{T}_i) - \hat{\mu}_i\right) \quad (16)$$

where $\hat{\alpha}$ and $\hat{\lambda}$ are the estimated shape and scale parameters respectively. The location parameter for the GP distribution represents the minimum activity period which can be estimated as in eq. (12). SU can acquire the knowledge about the minimum activity period through pilot signals broadcasting primary system's information locally or if the wireless technology of the primary system has a known standard.

The authors of [14] and [15] proposed a modified version of MoM where the mean and variance estimated from the reconstructed periods using MoM needed to be corrected with respect to the used estimation method as follows

$$\mathbb{E}(\tilde{T}_i) = \mathbb{E}(\hat{T}_i) + 2T_s \quad \text{for ES1} \quad (17.a)$$

$$\mathbb{E}(\tilde{T}_i) = \mathbb{E}(\hat{T}_i) - 2T_s \quad \text{for ES2} \quad (17.b)$$

$$\mathbb{E}(\tilde{T}_i) = \mathbb{E}(\hat{T}_i) \quad \text{for ES3} \quad (17.c)$$

$$\mathbb{V}(\tilde{T}_i) = \mathbb{V}(\hat{T}_i) - T_s^2/6 \quad \text{for all ES} \quad (17.d)$$

where $\mathbb{E}(\tilde{T}_i), \mathbb{V}(\tilde{T}_i)$ are the corrected mean and variance $\mathbb{E}(\hat{T}_i), \mathbb{V}(\hat{T}_i)$ are the estimated mean and variance.

After the estimation of the GP distribution parameters, the Cumulative Distribution Functions (CDFs) for the idle and busy periods of each channel are calculated according to eq. (6) using the estimated parameters. To study the effect of the compression ratio, sensing period, and sensing duration utilized by the compressive spectrum sensing scheme on the accuracy of the estimation of PU behavior, Kolmogorov-Smirnov (KS) distance $D_{ks}$ was used to compare the CDFs of both the actual and the estimated periods as follows

$$D_{ks} = \sup_{T_i} |F_{T_i}(T_i) - F_{\hat{T}_i}(\hat{T}_i)| \quad (18)$$

where $F_{T_i}(T_i)$ is the CDF of the actual activity periods and $F_{\hat{T}_i}(\hat{T}_i)$ is the CDF of the estimated activity periods.

Finally, we can estimate the duty cycle of the channels by calculating the mean of idle and busy periods using eq. (13) then the duty cycle can be calculated using eq. (9). To measure the accuracy of the estimated duty cycles we used Root Mean Square Error (RMSE) defined as

$$RMSE = \sqrt[2]{\frac{1}{N}\sum_{n=1}^{N}(\Psi_n - \widehat{\Psi}_n)^2} \qquad (19)$$

where $\Psi_n$ denotes the actual duty cycle value, $\widehat{\Psi}_n$ denotes the estimated duty cycle and $N$ is the number of channels.

## 4. SIMULATION AND EVALUATION

In our simulation we consider a single SU observing a frequency domain of 128 non-overlapping channels. Channels were divided into 7 groups and each group has a duty cycle $\psi$ where $\psi \in \Psi \approx \{0.01, 0.05, 0.1, 0.3, 0.5, 0.7, 0.9\}$. The number of channels per group was chosen to achieve a mean duty cycle $\overline{\Psi} = 0.1$ to represent a sparse frequency domain to be suitable for compressive spectrum sensing. In order to model the idle and busy periods of the primary system in each channel, periods were randomly drawn from the GP distribution where the parameters were chosen as in Table 1 to reproduce the channels' duty cycles by using eqs. (7) and (9). The parameters were selected as in the table to give PU behavior similar to that in [14], and to simulate the parameters of a real scenario like the parameters in [33]. The received signal at the SU is assumed to have a signal-to-noise ratio (SNR) of 30 dB. The Basis Pursuit Denoising (BPDN) algorithm is used by the SU to recover the spectrum which was sensed every $T_s$ seconds to produce a binary decision sequence for each channel, 1 if channel is occupied and 0 if channel is vacant.

The binary decisions are then used to estimate the idle and busy periods of the PU in each channel using ES3 method. After that, the GP distribution parameters and the distribution of the periods are estimated, and KS distance is calculated to evaluate the estimation error. In this paper, we investigated the results for duty cycles of 0.3, 0.5 and 0.7 as opposed to the results in [14] which only evaluated the results for only a duty cycle of 0.5. We did not investigate the results for too low or too high duty cycle values as channels with these duty cycles were assumed to be always idle or always busy, respectively. Our evaluation results show a similar KS distance performance for the estimation of the distribution of the periods as achieved in [14] and [15].

**Table 1: GP parameters used in the simulation**

| Duty Cycle (DC) | $\mu_1$ | $\lambda_1$ | $\alpha_1$ | $\mu_0$ | $\lambda_0$ | $\alpha_0$ |
|---|---|---|---|---|---|---|
| 0.29 | 0.5 | 0.35 | 0.0094 | 0.5 | 1.55 | 0.0134 |
| 0.5 | 0.5 | 0.5 | 0.010 | 0.5 | 0.5 | 0.010 |
| 0.71 | 0.5 | 1.28 | 0.011 | 0.5 | 0.22 | 0.0099 |

We firstly consider the effect of the compression ratio on the estimation of the distributions of the periods and the duty cycles. Fig. 2 (a) shows that the estimation of the distribution of busy periods is acceptable ($D_{ks} < 0.1$) if the number of measurements $M$ is in the order of $O(L_{max} \log(N/L_{max}))$ as $L_{max}$ denotes the maximum number of expected active PUs which can be seen at compression ratio $\geq 0.5$. At lower compression ratios (lower number of measurements) the accuracy of spectrum sensing decreases due to false alarms and miss detections which leads to a reduced estimation accuracy. A similar behavior for the estimation performance of idle periods distribution was observed. Fig. 2 (b) shows that at lower compression ratios the RMSE of duty cycles estimation is considerably high due to the imperfect spectrum sensing while at higher compression ratios the estimation of the duty cycles gets much better (RMSE $< 5 \times 10^{-3}$). Although a high compression ratio can provide accurate estimation, it costs CR units more sampling effort and more power consumption.

Secondly, we study the effect of the employed sensing period ($T_s$). As shown in Fig. 3 (a), at a compression ratio of 0.6 (high compression ratio was chosen so that estimation error is a function of sensing period only) the distribution of busy periods could be estimated with a mean KS distance $< 0.1$ (KS distance averaged over all channels in the same channel group) when $T_s$ is lower than the minimum period $\mu_i$ as proposed in [14], but at the cost of high power consumption. However, if the employed $T_s$ increased beyond the $\mu_i$ the accuracy of the estimation deteriorates due to skipping the periods having values greater than or equal to $\mu_i$ which results in overestimation of the mean value of idle and busy periods. So, the appropriate value of $T_s$ would be to equal $\mu_i$ which minimizes the sensing overhead and power consumption, and in the same time provides good estimation accuracy. Similar results were obtained for idle periods distribution estimation. Fig. 3 (b) shows that the estimation of the duty cycles is slightly affected by the employed $T_s$ as the duty cycle is calculated as the mean of busy periods divided by the sum of the mean of idle and busy periods leading to a slight effect as both mean values are overestimated together at higher $T_s$.

Finally, we evaluate the effect of the sensing duration. Fig. 4 (a) and (b) show that the estimation of the distribution of the activity periods as well as the estimation of the duty cycles can be enhanced if the spectrum sensing duration is increased as MMoM can give a better estimation when the sample size (the number of reconstructed periods) increases.

## 5. CONCLUSIONS

Compressive sensing is a suitable solution for wideband spectrum sensing. It allows CR units to sample the spectrum with sub-Nyquist rates instead of Nyquist sampling required by other spectrum sensing techniques. The knowledge of spectrum usage statistics can benefit SUs to perform spectrum sensing efficiently. It also helps in spectrum management and interference mitigation. Providing statistical information about PU activity to SUs through pilot signals and geo-located databases is not suitable for low resources CR units and for the instantaneous change of the spectrum occupancy. In this paper, we studied the effects of compressive spectrum sensing on the estimation of PU activity statistics. We showed that the distribution of idle and busy periods and the duty cycle of the channels can be estimated accurately using compressive spectrum sensing measurements while saving 40% of the Nyquist rate sampling needed by other sensing techniques. This can be achieved by choosing a sensing period below or equal to the minimum activity period and by selecting a compression ratio that provides enough measurements considering the maximum number of expected active PUs and a long sensing duration. These results can open the door for more efficient compressive spectrum sensing algorithms.

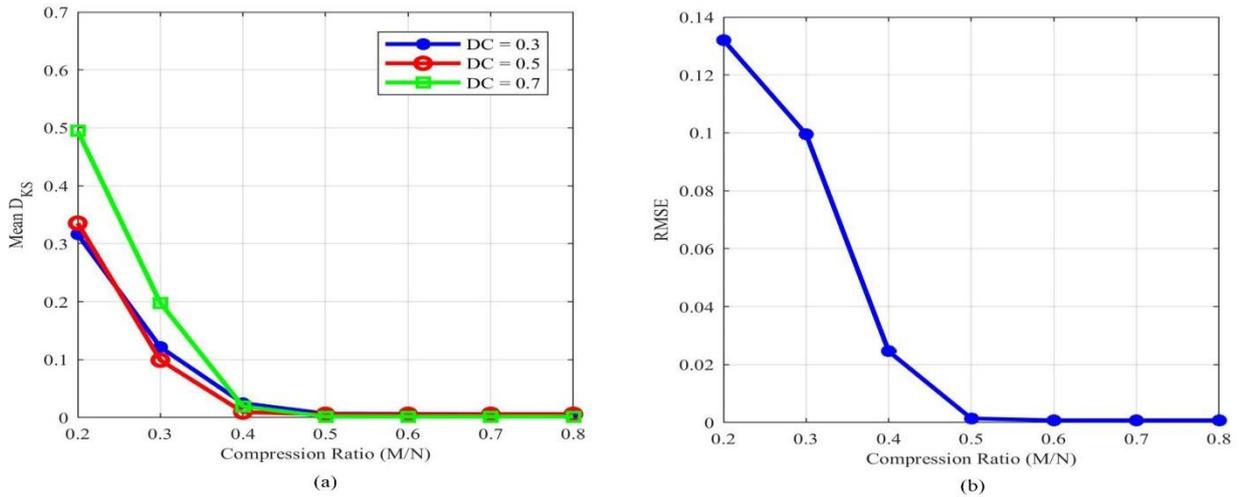

**Figure 2:** (a) Mean KS distance of busy periods distribution estimation at different compression ratio values, at $T_s = 0.5$ s, and (b) RMSE of duty cycle estimation at different compression ratio values, at $T_s = 0.5$ s, and sensing duration = 24 hours.

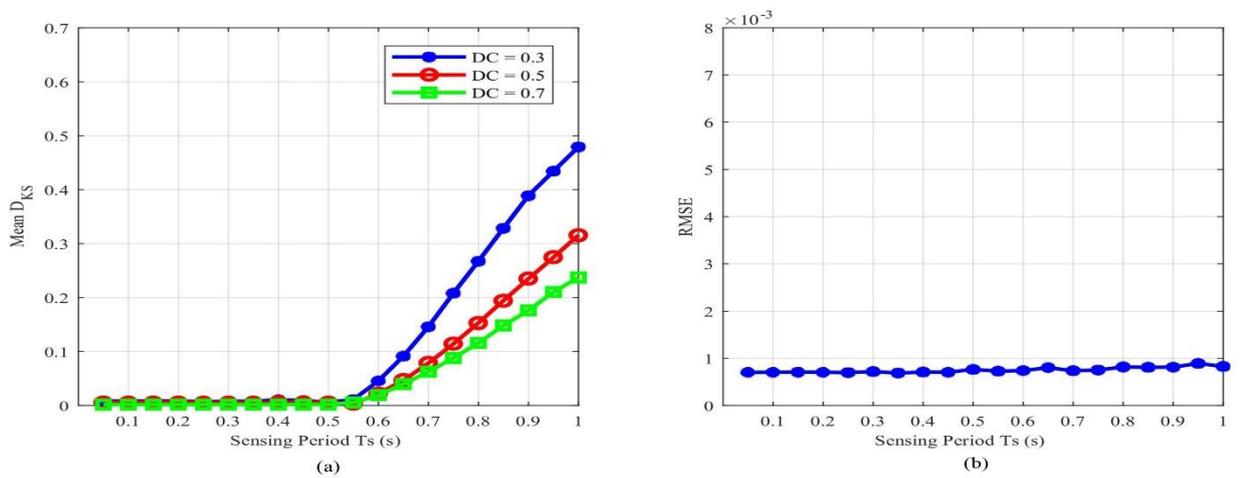

**Figure 3:** (a) Mean KS distance of busy periods distribution estimation at different $T_s$ values at a compression ratio of 0.6, and (b) RMSE of duty cycle estimation at different $T_s$ values at a compression ratio of 0.6, and sensing duration = 24 hours.

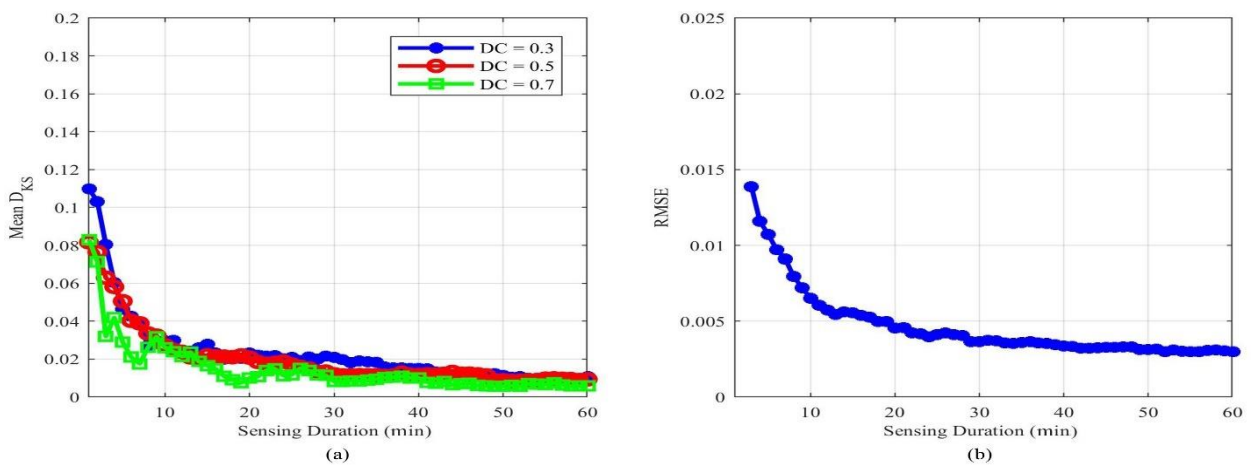

**Figure 4:** (a) Mean KS distance of busy periods distribution estimation at different sensing durations, at $T_s = 0.5$ s and a compression ratio of 0.6, and (b) RMSE of duty cycle estimation at different sensing durations, at $T_s = 0.5$ s and at a compression ratio of 0.6.

**Credit Authorship Contribution Statement:**
**Ahmed A. Tawfik**: Methodology, Writing Original Draft, Software.
**Mohamed F. Abdelkader**: Conceptualization, Review and Editing, Supervision.
**Sherif M. Abuelenin**: Supervision, Review and Editing.

**Declaration of competing Interest**
The authors declare that they have no known competing financial interests or personal relationships that could have appeared to influence the work reported in this paper.